\documentclass[preprint]{elsarticle}
\usepackage{graphicx}
\usepackage{amsmath}
\usepackage{tikz}
\usetikzlibrary{circuits.logic.US} 
\usepackage{mathtools}
\usepackage[font=footnotesize]{caption}
\usepackage{float}

\begin{document}
\title{Implementation of a custom time-domain firmware trigger for RADAR-based cosmic ray detection}
\author[ku]{S.~Prohira\corref{cor1}}
\ead{prohira@ku.edu}
\cortext[cor1]{Corresponding author}
\author[ku]{D.~Besson}%
\author[desy]{S.~Kunwar} 

\author[ku]{ K.~Ratzlaff}
\author[ku]{R.~Young}
\address[ku]{U. of Kansas, Lawrence, KS, U.S.A.}
\address[desy]{DESY, Zeuthen, Germany}

\begin{abstract}


Interest in Radio-based detection schemes for ultra-high energy cosmic rays (UHECR) has surged in recent years, owing to the potentially very low cost/detection ratio. The method of radio-frequency (RF) scatter has been proposed as potentially the most economical detection technology. Though the first dedicated experiment to employ this method, the Telescope Array RADAR experiment (TARA) reported no signal, efforts to develop more robust and sensitive trigger techniques continue. This paper details the development of a time-domain firmware trigger that exploits characteristics of the expected scattered signal from an UHECR extensive-air shower (EAS). The improved sensitivity of this trigger is discussed, as well as implementation in two separate field deployments from 2016-2017.

\end{abstract}
\maketitle
\section{Introduction} 

\begin{figure}[H]
\begin{centering}
\includegraphics[width=\textwidth]{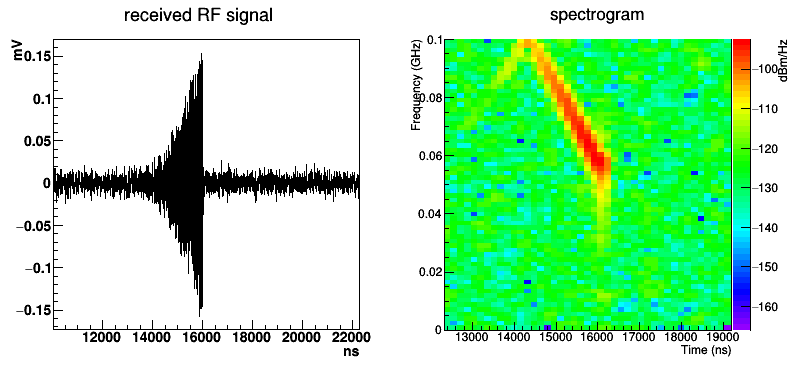}
\par\end{centering}
\caption{Expected radio reflection from a typical air shower. 54 MHz sounding frequency. Due to geometric effects, the amplitude increases as the shower progresses to the ground, while the frequency shift decreases. }
\label{sig}
\end{figure}

When an UHECR interacts with our atmosphere, a cascade of particles known as an extensive air shower (EAS) is initiated\cite{p_auger}. As the shower progresses to the ground, it ionizes a column of plasma through the atmosphere. For high enough incident energies, this plasma may be dense enough to reflect RF\cite{gorham}. The bi-static radio-scatter method of UHECR detection, as used by the TARA experiment \cite{tara}, involves a transmitter and a receiver, separated along some baseline with the transmitter illuminating a volume of the sky between them. An EAS within this volume will scatter the transmitter signal to the receiver. The shape, duration, and strength of this signal can then be used to determine the physical properties of the primary particle that caused the shower. A simulated received signal is given in Figure \ref{sig}. It is worth noting that this technique has been used for decades to detect and study meteors\cite{meteor1}.

Traditional cosmic ray detectors are ground-based\cite{auger}\cite{ta}, and detect the charged particles from EAS as they reach the ground. Since the flux of cosmic rays varies as $\sim E^{-3}$, the rate of the highest energy particles (O($10^{19}$)eV) is on the order of 1$km^{-1}century^{-1}$, making detection of the highest energy particles both cost and space prohibitive. The primary advantage of the radio-scatter scheme over traditional methods is that it offers, in principle, effective area coverage of many tens of $km^2$ with minimal apparatus\cite{dave_rf}. Though the TARA experiment reported no signal\cite{tara_limit}, we discuss here the firmware implementation of a different trigger method that may increase our sensitivity to EAS radio-scatter detection.

\section{Trigger}
\subsection{Theory}

The cold plasma produced by an EAS as it traverses the atmosphere is effectively stationary in 3-space, with ionization electrons having O(1-10 eV) energies. The plasma is not, however, stationary in 4-space, since the plasma lifetime in the troposphere and lower stratosphere is O(10ns)\cite{raizer}. Therefore, the moving shower front precedes a short-lived stationary plasma having an evolving number density. This unique physical phenomenon results in a scattered signal with a very unique signature; the phase relationships between different reflections from different parts of the shower combine coherently to result in a frequency-shifted return signal\cite{stasielak}, as illustrated in Figure \ref{sig}. This was exploited in the first iteration of the TARA Remote Stations (RS)\cite{sammy}, and is further exploited in the trigger described herein.

\begin{figure}[H]
\begin{centering}
\includegraphics[width=\textwidth/2]{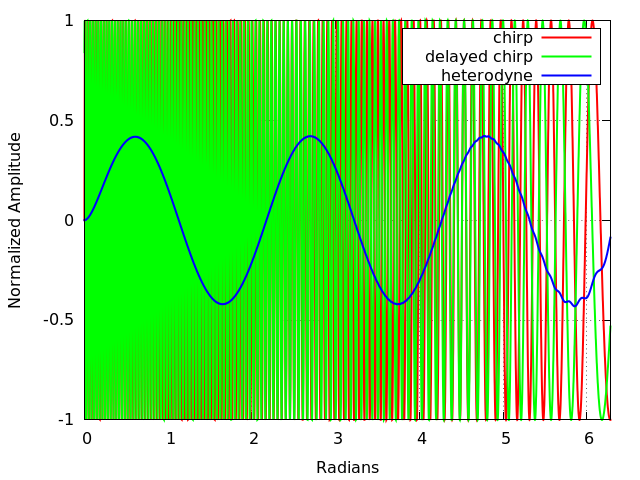}
\par\end{centering}
\caption{An example of the heterodyne method. A chirp with a starting angular frequency of 200Hz and a chirp rate $\kappa=15 Hz/s$ is mixed with a copy of itself, delayed by $\delta t = .1s$. The resultant monotone, $f_m$ is 3Hz. }
\label{chirp}
\end{figure}

``Heterodyning'' is the extraction of a modulation from a sinusoidal signal by mixing it with a second sinusoid, using the simple identity $2cos\theta cos\phi = cos(\theta-\phi)+cos(\theta+\phi)$. For FM radio, $\theta = (\omega_0 + \omega_{mod})t$, where $\omega_0$ is a carrier frequency and $\omega_{mod}$ is some audio-frequency modulation. Using $\phi = \omega_{lo}t$, where $\omega_{lo}=\omega_0$ is the local oscillator in a radio receiver, results in two heterodyne frequencies: the difference term (leaving only the modulation), and the sum term, which is up-shifted and easily filtered. The phase of a linear frequency-shifting signal may be written $\theta=\omega t + \kappa t^2$, where $\omega$ is the angular frequency and $\kappa$ is the rate of change of the frequency in units of $s^{-2}$. Setting $\phi$ to a time-delayed copy of $\theta$ with delay $\delta t$, the resultant difference heterodyne is a monotone frequency $f_m=2|\kappa|\delta t$. This is shown diagrammatically in Figure \ref{chirp}.

For an EAS we expect $\kappa$ of 1-3 Mhz/$\mu$s. Therefore, with a $\delta t$ of order 100ns, we expect monotones in the hundreds of kilohertz range. Post-mixing, we can envelope-detect the heterodyne output from the input linear chirp, and then trigger on these monotones.

 In the first revision of the RS \cite{sammy}, the delay was provided by a long cable, and the mixing and envelope detection was analog. This analog approach has been fully migrated to firmware to allow for a more robust trigger with greater versatility. 
 
 \subsection{Firmware implementation} 
 \subsubsection{Overview}
\begin{figure}[H]
\begin{centering}
\includegraphics[width=\textwidth]{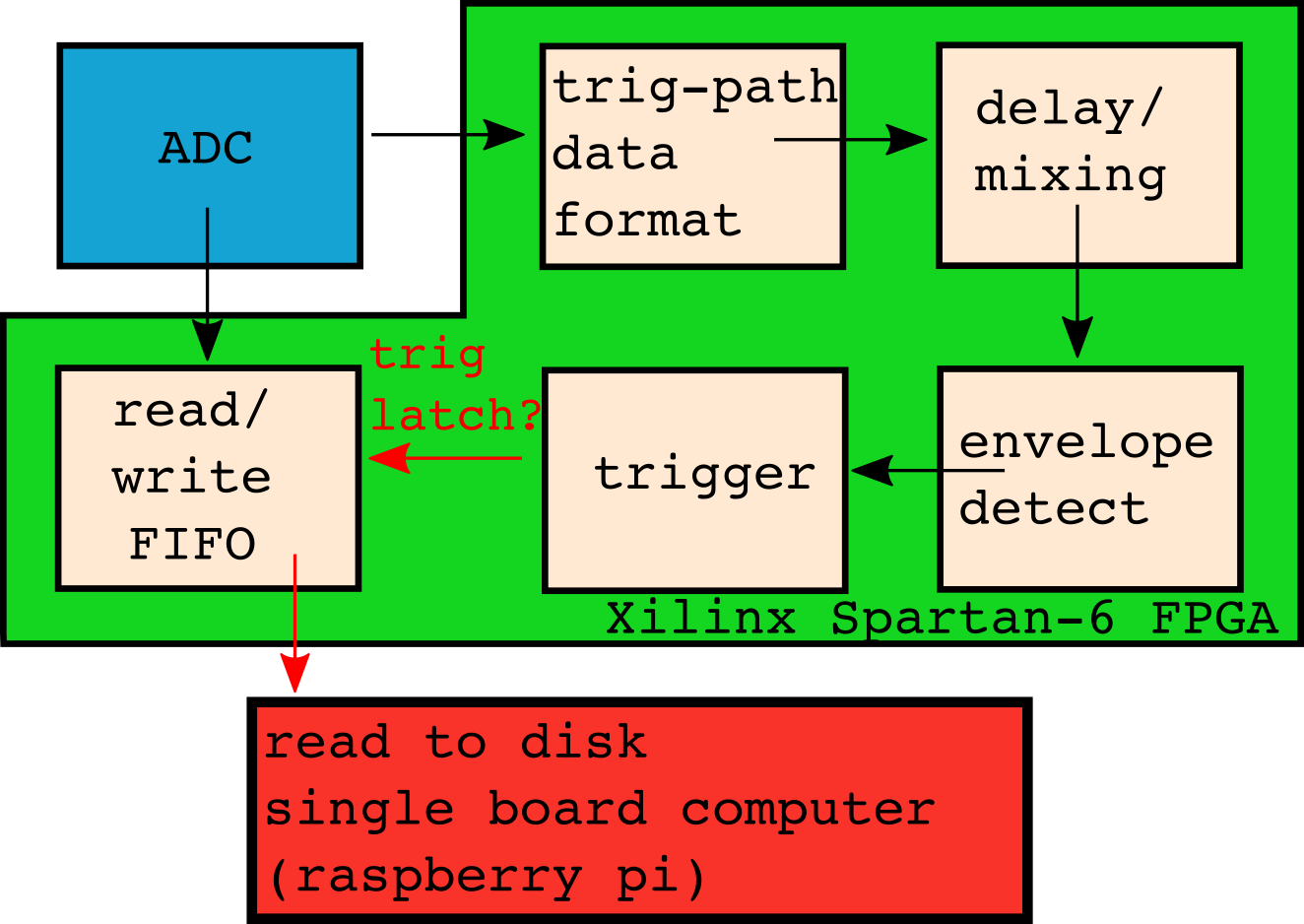}
\par\end{centering}
\caption{Block diagram of the RS trigger path. Upon satisfaction of trigger logic, the write-data FIFO is read out and written to disk. }
\label{signalchain}
\end{figure}

The heterodyne-output trigger is now a series of modules on a Xilinx Spartan-6 FPGA\cite{xilinx}, written in VHDL. A block diagram of the signal chain is given in Figure \ref{signalchain}. An Analog Devices AD9634 analog-to-digital converter (ADC) clocked at 250 MS/s digitizes our incoming signal with 12-bit resolution. This signal is then split, with one half routed into a 16,384 word first-in, first-out (FIFO) buffer, and the other half routed to the trigger path. The read/write FIFO is written to disk when a trigger is registered. 

The trigger path begins with a formatting module that takes the incoming ADC and formats it into a string of signed 12 bit words. This formatted data is then sent to a 32 bit FIFO which serves as our mixer. We set a read point (25 samples x .4 ns per sample = 100 ns) into this FIFO as our delayed signal. The mixer outputs a normalized sample-by-sample product of the two input samples, direct and delayed. This output is then sent to the envelope-detection module, which is primarily a single-pole, infinite-impulse-response (IIR) low-pass filter. This filter configuration was selected due to the rapid response times achievable with an IIR. This envelope is then sent to a 3-part trigger module. 
Details of the individual portions of the trigger path are given below, and a plot of Xilinx Chipscope on-chip traces is shown in Figure \ref{trig}, with the various signals labeled. 

\begin{figure}[H]
\begin{centering}
\includegraphics[width=\textwidth]{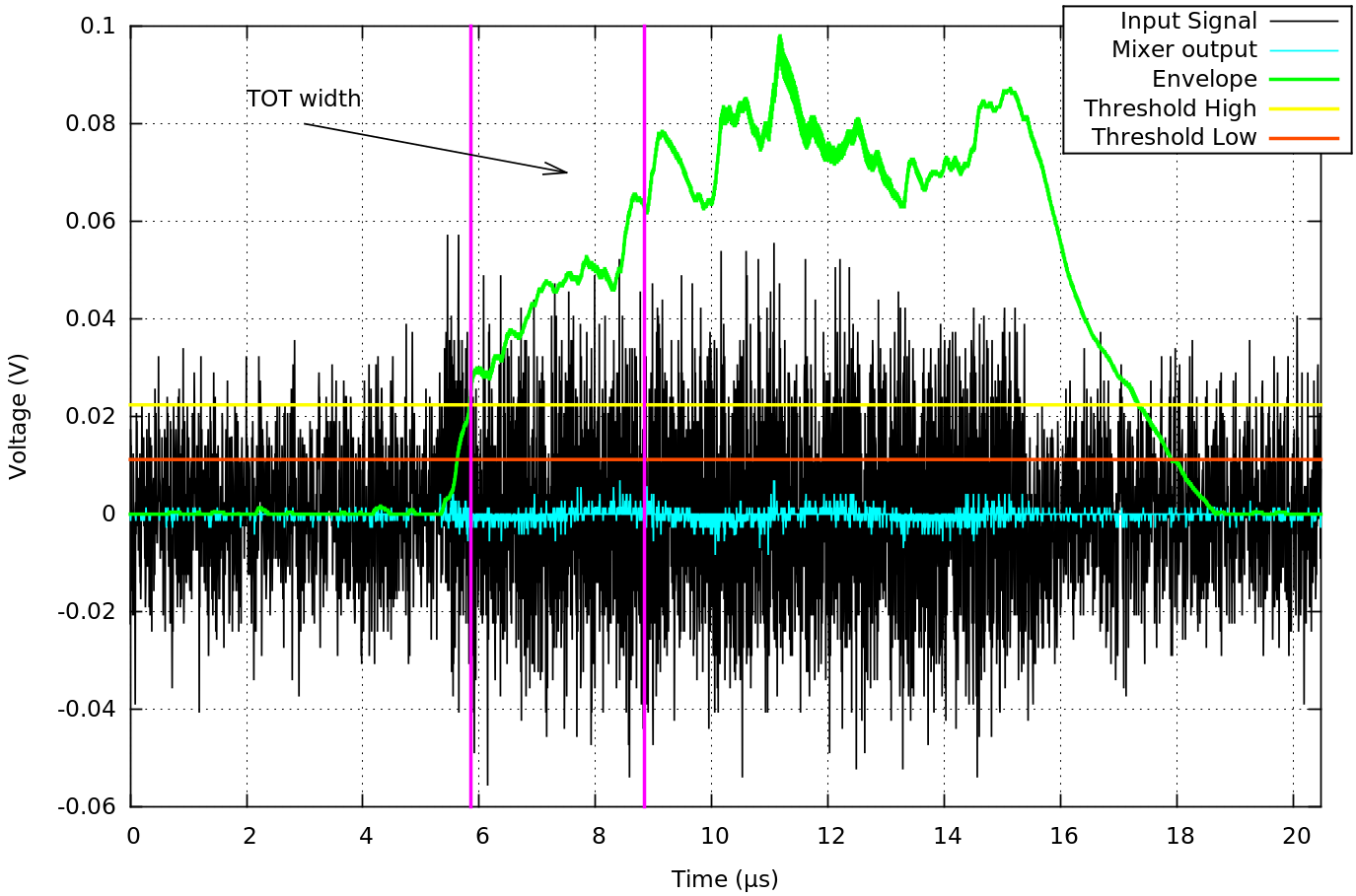}
\par\end{centering}
\caption{An example trigger, showing CHIPSCOPE on-chip signals. A low SNR linear chirp is embedded within noise and sent through the trigger path, resulting in the very high SNR output trace (green). }
\label{trig}
\end{figure}
\subsubsection{Formatting}
The data from the ADC has 12 bit resolution, sent to the FPGA as the 6 central bits of two consecutive 8-bit words. Though the raw data is stored this way in the write-buffer FIFO for continutity, it is an inconvenient format for mathematical operations. Therefore the formatting module strips the MSB and LSB from each word and concatenates the two consecutive words that make up 1 sample. These are then passed as signed 12-bit vectors through the remainder of the trigger path.

\subsubsection{Mixer}
The mixer module is a simple dual-port RAM FIFO implemented with the Xilinx CORE generator, with a write depth of 30 samples. This allows us to change the delay between direct and delayed signals in the heterodyne as needed. The frequency of the heterodyne monotone is proportional to the delay time, by $f_m=2|\kappa|\delta_t=2|\kappa|n T$, where n is the number of delay samples and $T$ is the sampling period. Each input sample is mixed with the delayed sample specified by the read point, which provides the heterodyne data. The output is normalized to prevent overflow further down the trigger path. 
\subsubsection{Filtering: Envelope detection}
The mixer output is then sent to the envelope-detector module. In hardware, a simple envelope detector is a series rectifier and resistor followed by a capacitor to ground. In firmware, we can simply square the input values and then low-pass filter them. We chose a custom single-pole IIR because the calculation only delays the input by 2 samples, as opposed to a finite-impulse-response filter, which typically need hundreds of samples to achieve similar filtering capabilities\cite{lyons}. The simple differential equation governing the low-pass filter described above, with associated resistor and capacitor values R and C is
\begin{equation}
\frac{v_i-v_o}{R}=C\frac{dv_o}{dt}
\end{equation}
Where subscripts designate the input and output voltages, and the difference is the voltage drop across the resistor. With the assistance of Laplace transforms, this can be translated into a time-domain transfer function $H(t)=x(t)/y(t)$, where $x, y$ are the input and output circuit voltages as a function of time,
\begin{equation}
H(t)=\omega_{rc}e^{-\omega_{rc}t}
\end{equation}
Where $\omega_{rc} = 1/RC$. We can then discretize this signal for our digital application using a discrete version of the Laplace transform, and with the substitution $y(t)\rightarrow y[n]$ arrive at an expression that can be translated into VHDL,
\begin{equation}
y[n]=\omega_{rc}x[n]+e^{-\omega_{rc}T}y[n-1]
\end{equation}

where $n$ is the current sample and $T$ is the sampling period. Motivation for the term ``infinite-impulse response'' is evident in this last expression-each filter output value is a function of both the incoming data sample $x[n]$ and the last filter output sample $y[n-1]$. Because of this recursion, the output values theoretically only ever approach zero asymptotically and hence persist at a non-zero value infinitely (though zero values are achieved in practice due to limitations of resolution). 

Though a sharper cutoff is possible with higher-order filters, each extra pole in an IIR introduces instability and delay. We found empirically that the single-pole filter offered the best mix of filtration and stability. Therefore, our module implements the above single-pole IIR, with coefficients chosen for a -3dB point at 300KHz. The input stream from the mixer is re-sampled at 10MHz, squared, fed through the filter, and then a 4 sample running-average is used to smooth the output, which is then sent to the trigger module. 

\subsubsection{Trigger}
\begin{figure}[h]
\begin{centering}
\begin{tikzpicture}[circuit logic US, every circuit symbol/.style={thick}]
	\node[and gate] (and1) at (0,1) {};
	\node[not gate] (not1) at (0,-.1) {};
	\node[and gate] (and2) at (1.5,0) {};
	\draw [black, thick] (and1.output) -- ++(right:.25)--++(down:.9) --(and2.input 1);
	\draw [black, thick] (not1.output)-- (and2.input 2);
	\draw [black,thick] (and1.input 1) -- ++(left:1) node [left]{Edge};
	\draw [black,thick] (and1.input 2) -- ++(left:.5)--++(down:.4)--++(left:.5) node [left]{TOT};
	\draw [black, thick] (not1.input) --++(left:1.1) node [left]{Veto};
	\draw [black, thick] (and2.output) --++(right:1) node [black,right]{Trigger out};
\end{tikzpicture}
\end{centering}
\caption{Trigger logic circuit diagram.}\label{ckt}
\end{figure}
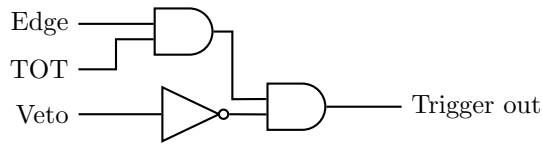

The 3-fold trigger logic uses the simple circuit shown in Figure \ref{ckt}. There is a threshold requirement, a time-over-threshold requirement, and a transient veto module. The threshold is a simple two-fold edge threshold. The incoming data must rise above a high threshold and then remain above a low threshold to satisfy the edge trigger logic. The TOT logic goes high once the input stream has satisfied the edge trigger logic for a specified amount of time, and thereby suppresses short-duration signals. The third parameter is a transient veto. In this module, the time derivative of the incoming envelope is monitored, and if it rises above a certain threshold, this logic goes high and vetoes the remaining trigger logic. This is an important addition to the trigger because very short, high-amplitude transients are highly broadband, and thus can thwart both the heterodyne and the low-pass filter. In addition, since the amplitude envelope of the expected EAS reflection steadily increases as the shower progresses, as evidenced in the simulated signal of Figure \ref{sig}, the slope of the envelope for a true signal should be very small. The three parameters-edge threshold, TOT width, and transient-veto threshold are set at runtime via commands sent to the FPGA over SPI. Examples of trigger logic are given in Figure \ref{trig}.

\begin{figure}[H]
\begin{centering}
\includegraphics[width=\textwidth]{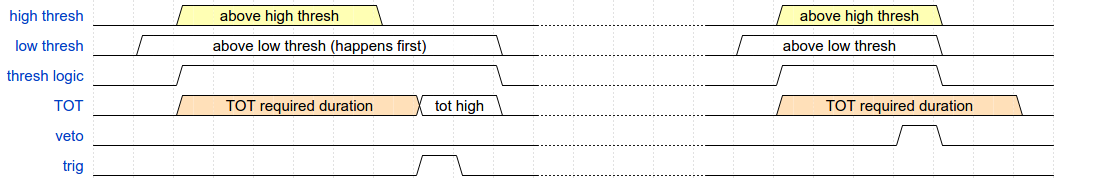}
\par\end{centering}
\caption{example timing diagrams of the trigger system. Left: The trigger envelope rises above low threshold then high threshold. after some amount of time, the time-over-threshold is satisfied, and the trigger goes high. the veto in this case stays low, indicating that there is not a transient present. Right: The trigger envelope rises above the low threshold and also the high threshold, but the veto goes high, which sends the threshold and time-over-threshold logic low, and so the trigger never goes high.}
\label{trig}
\end{figure}
In addition to negating the trigger, the veto module also communicates with the envelope module to zero the IIR filter, thereby counteracting one of the pitfalls of the IIR design. Due to recursion, the impulse response to high-amplitude signals requires many samples to relax to unobservable levels, so that if a sufficiently-high amplitude impulse arrives at the filter input and sends the transient veto high, the decay of the envelope could satisfy both the edge and TOT requirements even once the veto goes low, resulting in a false trigger. Therefore, when the veto goes high, it signals the envelope-detect module to zero the filter values, killing the impulse decay, and resetting the filter. Due to the simplicity of the IIR, the filtration/envelope detection recovers in 2 samples after this veto, resulting in a minimum of efficiency loss. 

\begin{figure}[H]
\begin{centering}
\includegraphics[width=\textwidth]{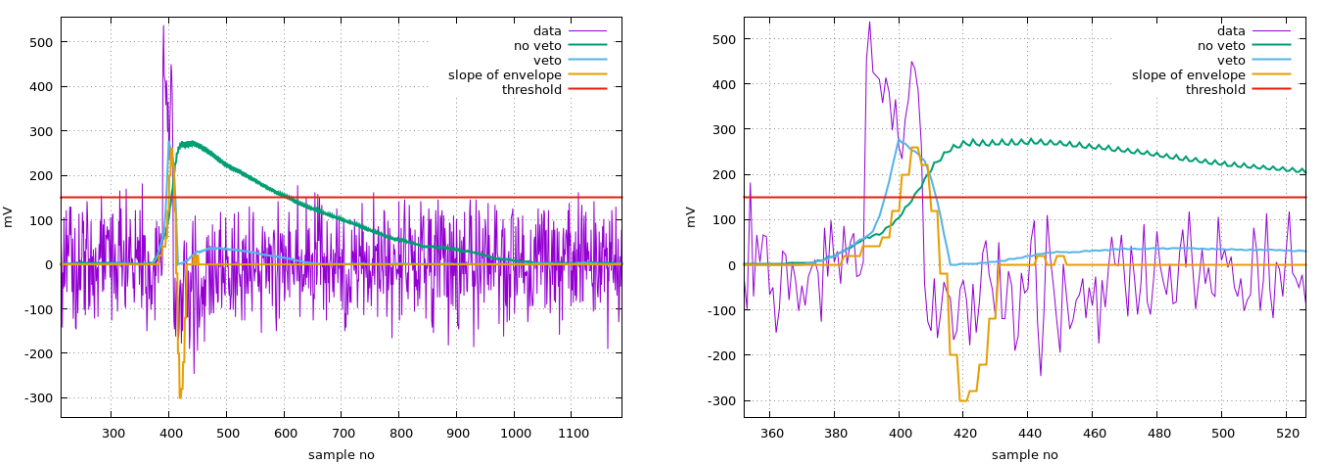}
\par\end{centering}
\caption{Operation of the trigger veto system. Due to the broadband nature of transients, high-amplitude spikes will trip the envelope detector, which rises above the threshold and satisfies the TOT width (green online). With the trigger veto, which monitors the slope of the envelope rise, the envelope is truncated before the TOT can be satisfied.}
\label{veto}
\end{figure}

A plot showing the function of the transient veto module is given in Figure \ref{veto}. High amplitude transients, which would trip a simple edge trigger, are killed by the veto so that they cannot satisfy the TOT requirement. 

If a trigger is registered, the read/write FIFO is then latched and written to disk, which in this case is the on-board flash memory of a Raspberry Pi single board computer (SBC). The SBC communicates with the FPGA via an SPI interface, controlled using the open source WiringPi SPI library for C. 

\section{Performance}
A first deployment of this system, without the transient veto and running at 200MHz, was part of the TARA experiment from Feb-April 2016. As of this writing, a re-depolyment of the full system with veto, just north of Lawrence, KS at the Kansas Biological Survey KU Field Station, is underway. The following section represents data taken during the TARA deployment. 
\subsection{System Sensitivity}
\begin{figure}[H]
\begin{centering}
\includegraphics[width=\textwidth]{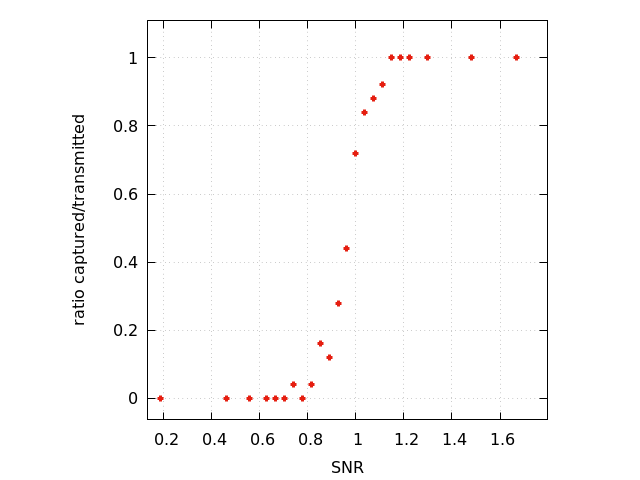}
\par\end{centering}
\caption{Ratio of captured pulses to transmitted pulses as a function of the SNR of the input chirp, for a chirp slope of -1.5 MHz/$\mu$s. The trigger turns on at SNR $<$ 1. }
\label{trig_eff}
\end{figure}

The trigger is sensitive to signals at low SNR. The trigger efficiency for a single chirp slope is given as a function of SNR in Figure \ref{trig_eff}. As evidenced by the figure, sensitivity begins below SNR of 1.

\subsubsection{In-Field Calibration}
\begin{figure}[H]
\begin{centering}
\includegraphics[width=\textwidth]{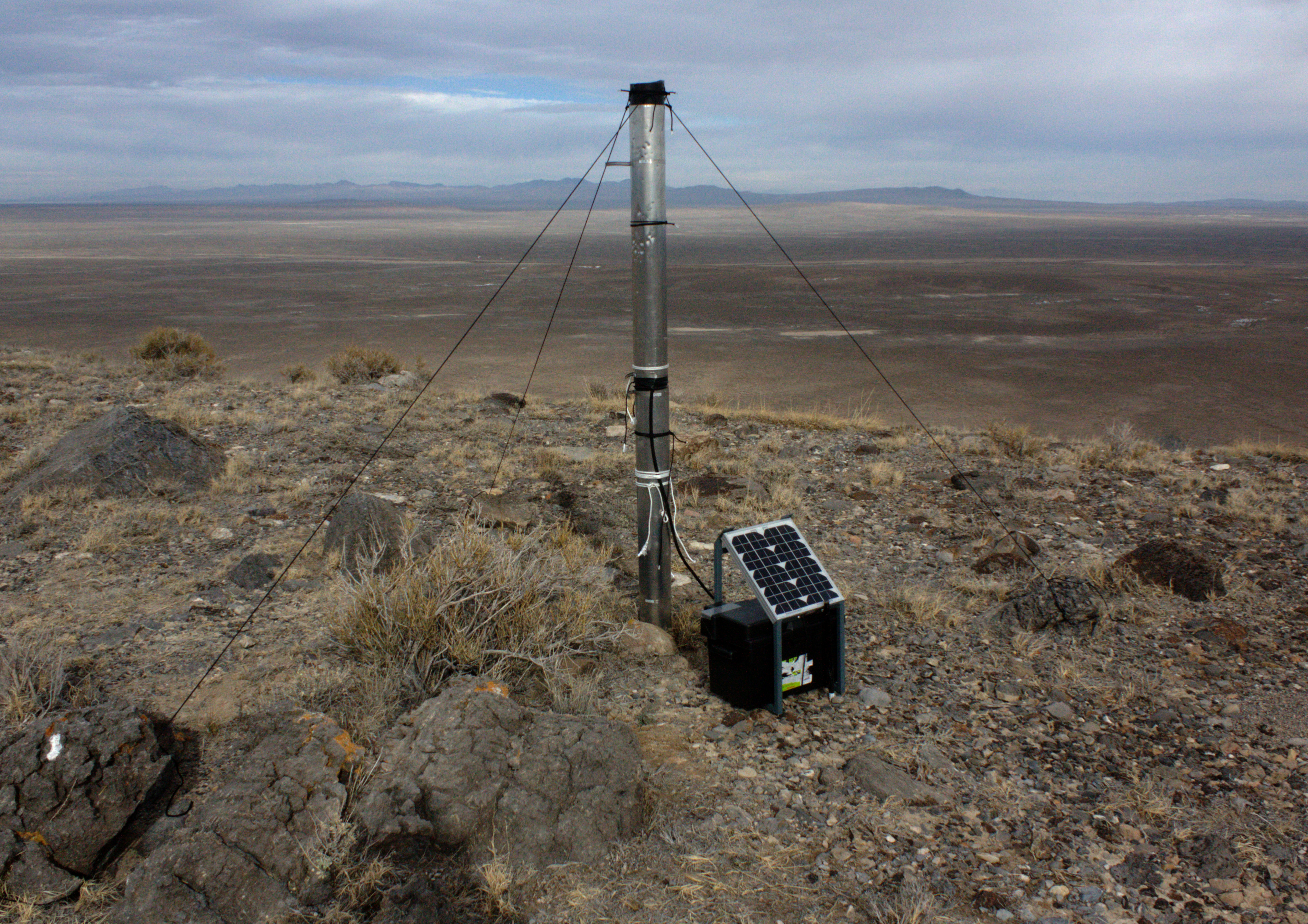}
\par\end{centering}
\caption{The chirp calibration unit deployed on Long Ridge, UT.}
\label{ccu}
\end{figure}
\begin{figure}[H]
\begin{centering}
\includegraphics[width=\textwidth]{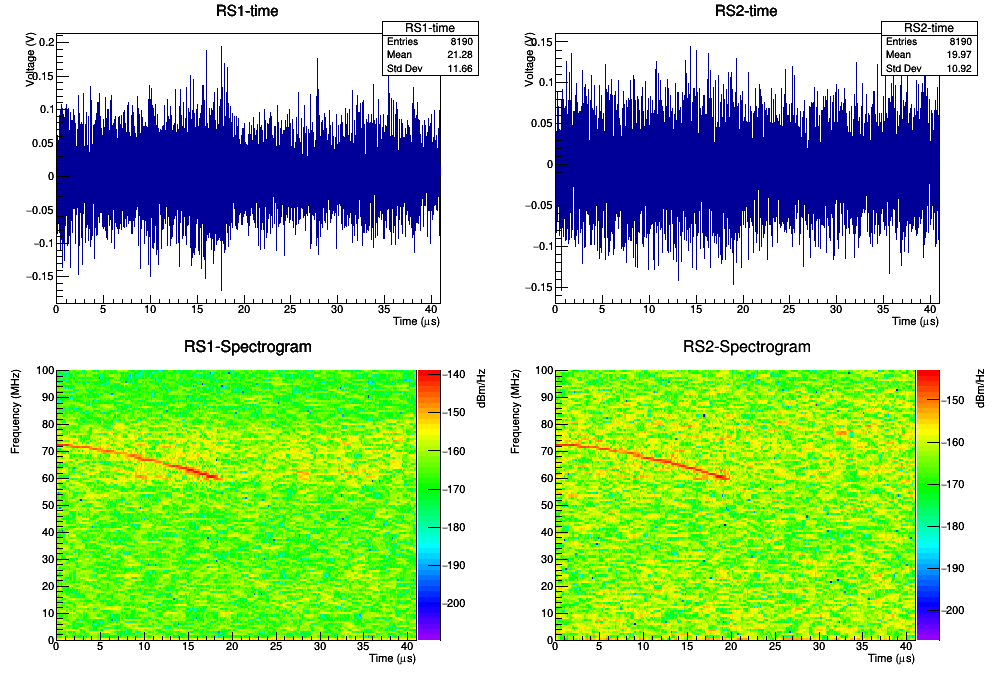}
\par\end{centering}
\caption{An example CCU pulse captured coincidently in RS1 and RS2 during deployment on Long Ridge, UT. }
\label{ccu_ex}
\end{figure}

Deployed along with the RS was a system called the Chirp Calibration Unit (CCU), shown in Figure \ref{ccu}, which output a periodic chirp signal to calibrate the RS systems. The CCU is a custom board featuring a voltage controlled oscillator (VCO) governed by an ATMEL ATMEGA328 microcontroller, designed and built at the Instrumentation Design laboratory at KU. The chirp slope and duration of the VCO are set in the microcontroller firmware. The output of this board is attenuated to the desired amplitude and coupled to a custom `fat' dipole, with a resonance at 70MHz and a sub-3 VSWR bandwidth of $\pm$20 MHz for full coverage of our expected signal region. An in-field CCU chirp captured simultaneously in RS1 and RS2 is given in  Figure \ref{ccu_ex}, showing that the full system, as deployed, is sensitive to chirps at SNR $\sim$1. 

\subsubsection{Galactic background}
\begin{figure}[H]
\begin{centering}
\includegraphics[width=\textwidth]{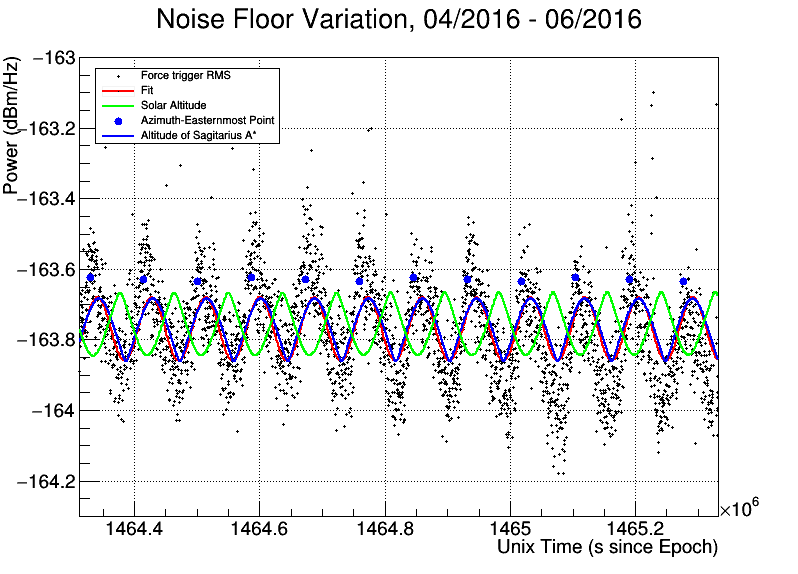}
\par\end{centering}
\caption{RMS power of forced triggers for April-June of 2016, zoomed in to a 12 day region. A sine-wave fit to the data matches very well with the altitude of the galactic center with respect to the boresight pointing of the antenna. Variation due to the solar altitude (green) is in poor agreement with the fit, suggesting true sensitivity to the galactic background. }
\label{galaxy}
\end{figure}

Another important metric is the measurement of the ambient background temperature. In principle, the ultimate noise floor to which a small-signal RF experiment should be sensitive to is the CMB. However, local galactic sources make it impossible to achieve this noise floor, as they contribute to the background temperature. One way to monitor the sensitivity of the system to galactic backgrounds is to take forced trigger snapshots periodically, and monitor the change in ambient background. For the deployment of RS rev. 2, these snapshots were taken at 6 minute intervals. Figure  \ref{galaxy} shows a clear diurnal variation that correlates poorly with the solar altitude, but quite well with the altitude and azimuth of the galactic center with respect to the boresight pointing of the RS antenna. Such a study is essential to verifying that we are sensitive to the lowest possible signal levels, and not simply amplifying our own system noise.

\subsection{Pointing}

The RS employ a timing system that affords O(10) ns timing resolution per station. An iLotus M12M GPS chip provides a 100 pulse-per-second (PPS) line which is read by the FPGA, controlling 2 counters. A slow counter simply counts the number of 100PPS pulses since the top of the GPS second, which is indicated by a wider than usual pulse. A fast counter counts the number of system clock cycles between successive 100PPS pulses. These two counts are latched when a trigger is registered, and are written with the ADC data. 

\begin{figure}[H]
\begin{centering}
\includegraphics[width=\textwidth]{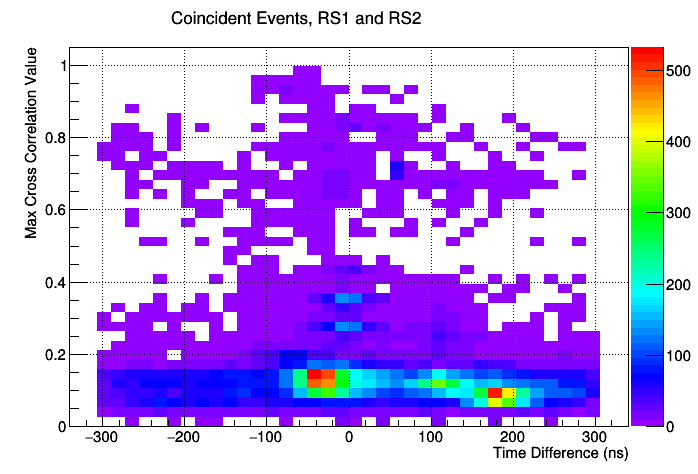}
\par\end{centering}
\caption{maximum cross-correlation value of coincident events between RS1 and RS2, as a function of the difference between the trigger timestamps. The causal window for events is $\pm$200 ns, since the stations were placed roughly 70 meters apart. Events out near $\pm$200ns are aligned along the axis connecting the two antennae, while events near 0, which arrive at the antennas simultaneously, come from sources equidistant to both stations. The bright spots at 0ns and cross-correlation values of .25-.4 are the CCU for varying levels of SNR. Bright hot spots at low cross correlation values correspond to man-made structures in the vicinity.}
\label{xcor}
\end{figure}

This high-resolution timing allows for the two stations to `point' at the sources of signals. Naturally two stations cannot isolate a point in 3 space, but the knowledge that most anthropogenic backgrounds are restricted to the ground plane effectively allows us to use pointing to isolate noise sources. Figure \ref{xcor} shows the maximum cross-correlation value of coincident events captured by RS1 and RS2 as a function of the time difference of the trigger timestamps (RS1-RS2). The trigger point slewing inherent to time domain threshold triggers has been corrected for algorithmically. Figure \ref{xcor} clearly shows noise sources, as well as the CCU in 2 different amplitude configurations, as hot spots in the plot. Events which do not obviously cluster with others, are in the largely forward region (small $\Delta$t), and which have a high cross-correlation value will be the events of interest in the final analysis.

\subsection{Durability}
The stations were deployed in the Utah desert from from February through December of 2016, though they only took active data through August. The TARA transmitter ran from Feb-May, and after that the RS only took forced-trigger data, from which the above galactic background plots were made. This means that the stations ran in sub-zero freezing conditions during winter, and also through average summer temperatures of 35 C. The CCU, as well as both stations, returned after this deployment in perfect working order.

\section{Summary}
The RS are a custom-built, autonomous detector system for UHECR physics. The system is designed to trigger on the reflected RF signal itself, and can trigger at low SNR. The current trigger, running at 250MHz and including the transient veto, is deployed at the Kansas Biological Survey Field Station, north of Lawrence, KS. This system uses FM radio transmitters as the sounding RF, allowing for illumination of a greater volume with greater power. This run is scheduled to last through autumn 2017.

\section{Acknowledgements}
This work was supported by the WM Keck Foundation and an award from the Kansas Biological Survey KU Field Station. The authors would like to thank the TARA collaboration and the TA collaboration for their cooperation and support.

\bibliographystyle{elsarticle-num}

\bibliography{/home/natas/Documents/physics/tex/bib}

\end{document}